\documentclass[preprint,12pt,authoryear]{elsarticle}
\usepackage{graphicx}
\usepackage{epstopdf, epsfig}
\usepackage{subcaption}
\usepackage[section]{placeins}
\usepackage[export]{adjustbox}
\usepackage{gensymb}
\usepackage{amsmath}
\usepackage{dirtytalk}
\usepackage[section]{placeins}
\usepackage{needspace}
\usepackage{color}
\usepackage[utf8]{inputenc}
\usepackage[T1]{fontenc}

\usepackage{siunitx}

\usepackage{amssymb}

\journal{International Journal of Multiphase Flow}

\begin{document}

\begin{frontmatter}



\title{Mixing and solvent exchange near the turbulent/non-turbulent interface in a quasi-2D jet}

\affiliation[inst1]{organization={Physics of Fluids Group and Max-Planck, Center for Complex Fluid Dynamics, Faculty of Science and Technology, J.M. Burgers Center for Fluid Dynamics, University of Twente},
            addressline={}, 
            city={Enschede},
            postcode={7500 AE}, 
            state={},
            country={Netherlands}}

\affiliation[inst2]{organization={Max Planck Institute for Dynamics and Self-Organization},
            addressline={Am Fa\ss berg 17}, 
            city={G\"{o}ttingen},
            postcode={37077}, 
            state={},
            country={Germany}}

\author[inst1]{You-An Lee}
\author[inst1]{Sander G. Huisman}
\author[inst1,inst2]{Detlef Lohse}

\begin{abstract}
We inject with jet mixtures of ethanol and dissolved anise oil upward into quiescent water with jet Reynolds numbers, $500<Re_0<810$. Nucleation of oil droplets, also known as the ouzo effect, follows from the entrainment and mixing with the ambient water, in which the oil has much lower oil solubility than the initial jet fluid. We experimentally investigate the local concentration during solvent exchange near the turbulent/non-turbulent interface (TNTI) in the quasi-two-dimensional turbulent jet. Using a light attenuation technique, we measure the concentration fields for the solvent exchange case (the nucleated oil droplets) and the reference dye case (passive scalar). 

Combining conventional and conditional mean profiles, we show that the nucleation of oil droplets is initiated near the TNTI upstream, and penetrates into the turbulent region as the jet travels downstream. Persistent nucleation not only sustains the scalar dissipation rate, but also leads to enhanced temporal fluctuations of the concentrations till the downstream (high) position. The probability density functions (PDFs) of the concentration exhibit pronounced bimodal shapes near the TNTI and positively-skewed curves toward the centerline, which also suggest that the oil droplets nucleate across the jet upon mixing.

In addition to the qualitative investigation, we also describe the process more quantitatively. Based on the idea that the concentration field of the nucleated oil micro-droplets is linked to the ethanol concentration field running in the background, we estimate the nucleation rate near the TNTI using a control volume approach. We model the concentration field of the nucleated oil using that of the passive scalar and the ternary phase diagram of a water, ethanol, and anise oil mixture.

This study extends our previous work on the mean field, revealing more details of the turbulent statistics induced by solvent exchange. The findings shed new light on the interplay between mixing and nucleation in a quasi-2D turbulent jet. We also provide the first modelling of solvent exchange in turbulent flows with a simple model based on ethanol concentration field and the phase diagram.
\end{abstract}

\begin{graphicalabstract}
\includegraphics[scale=1]{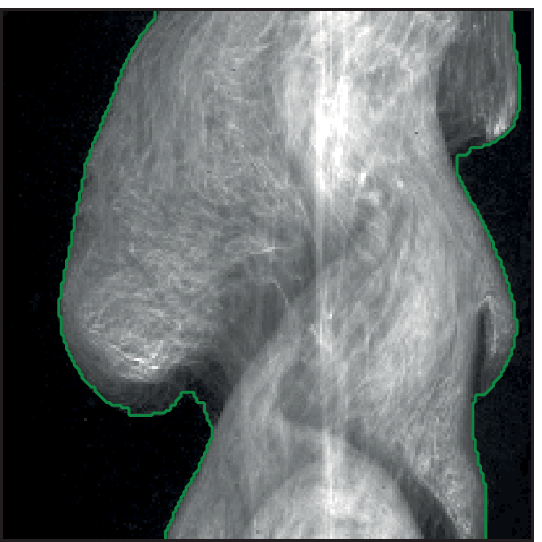}
\end{graphicalabstract}

\begin{highlights}
\item Mixing near the turbulent/non-turbulent interface induces solvent exchange and initiates droplet nucleation.
\item Concentration field in the opaque flow is measured using a light attenuation technique. 
\item Continuous solvent exchange leads to prolonged temporal and spatial concentration fluctuations across the entire turbulent jet.
\end{highlights}


\end{frontmatter}


\section{Introduction}

Turbulent mixing has been of central interest in the turbulence community for decades; it is a fundamental and essential topic for turbulence research \citep{Rothstein1999, Dimotakis2005, Villermaux2019}. Built upon the knowledge of passive scalar mixing, more complicated processes related to mixing also attract tremendous attention due to their relevance to various industrial applications and natural phenomena. We refer to these processes as physicochemical hydrodynamics, which can be broadly categorized into chemical reactions, in particular combustion \citep{Peters2000, Pitsch2006,Bisetti2012, Attili2014a}, and thermodynamic phase transitions \citep{Lesniewski1998, Murfield2013, Rivas2016, Lohse2020}. While the mechanisms behind these physicochemical hydrodynamical processes are very different, they all involve the conversion from one substance or phase to another due to turbulent mixing. 

For reactive flows, the competition between the timescale of mixing at small scales and that of the chemical reaction determines the rate-limit of the process and the resulting velocity and scalar fields, which is characterized by the Damk$\ddot{\text{o}}$hler number, the ratio between the hydrodynamics and reaction timescales. \citet{Karasso1996} and \citet{Mingotti2019b} studied chemical reactions in turbulent mixing layers and turbulent plumes, respectively. They both indicated that mixing is the rate-limiting process for fast reactions, i.e., those with high Damk$\ddot{\text{o}}$hler number. On the other hand, for slow reactions, the reactant and the product scalar field depend mainly on the chemical kinetics of the involved reaction. To characterize turbulent mixing, in addition to the scalar field itself, the scalar dissipation rate, the spatial gradient of the scalar field, gives additional insight. This quantity is key to the precise modelling of turbulent combustion. \citet{Bilger2004} suggested that the scalar dissipation rate is proportional to the reaction rate. Furthermore, \citet{Mellado2009} and \citet{Sutton2013} used the dissipation rate conditioned on the mixture fraction, where the mixture fraction measures the mass fraction of the fuel stream in the fuel/air mixture. They showed a non-monotonic dependence and a local optimum of the dissipation rate in the mixture fraction parameter space.   

While the aforementioned studies focused on the scalar fields in a continuous phase, physicochemical hydrodynamics following turbulent mixing can also lead to precipitation of a dissolved component into a dispersed phase, which undergoes oversaturation, droplet nucleation, droplet growth, and sometimes also shrinkage due to evaporation \citep{lance2014,Li2023} in the turbulent flow. Such precipitation processes include aerosol formation through condensation \citep{Lesniewski1998,Zhou2014, Ng2021, Li2023}, soot formation in turbulent combustion \citep{Bisetti2012,Attili2014a}, particle formation by chemical reaction \citep{Schikarski2022, Tang2022}, and droplet formation upon solvent exchange \citep{Lee2022}. \citet{Lesniewski1998} experimentally studied dibutyl-phthalate (DBP) aerosol formation in a turbulent jet flow, indicating that with high DBP vapor concentration, the droplets nucleate both within the initial shear layer and in the downstream turbulent region. \citet{Zhou2014} also investigated the formation of the DBP particles, in a mixing layer using a numerical approach. They showed that nucleation and growth of the particles are preferentially located at different sides of the mixing layer. \citet{Murfield2013} focused on the modelling of nucleation in general turbulent flows. They confirmed the experimental finding of \citet{Lesniewski1998} that nucleation of the DBP particles was initiated both in the initial shear layer and by the large-scale eddies. Combining experiments and numerical modelling, \citet{Schikarski2022} investigated the precipitation of ibuprofen particles in a T-mixer upon mixing and reaction, aiming at pharmaceutical applications. They showed the effect of the Damk$\ddot{\text{o}}$hler number on the particle size distribution, and demonstrated that their numerical framework correctly predict the ibuprofen particle size distribution. 

Solvent exchange with subsequent precipitation, also known as the ouzo effect, is a substance-extraction process utilizing a solubility constraint in a multi-component solution. The solution consists of a good solvent, a solute, and a poor solvent. Mixing the good solvent with a sufficient amount of the poor solvent decreases the solubility of the solute, leading to saturation and nucleation, which will either be collected or dispersed in the flow. While the physicochemical hydrodynamics of the ouzo effect has been extensively studied in the laminar regime in microfluidic setups \citep{Hajian2015, Zhang2015, Li2021}, it remains almost unexplored in the turbulent regime. In our previous work \citep{Lee2022}, we studied the concentration field of nucleated oil in an axisymmetric turbulent jet. Implementing a light attenuation technique, we successfully measured the mean structure of the field. The results revealed the effects of continuous nucleation in the jet, including reduced centerline decay rate, larger jet width, and monotonically increasing mass flow rate with streamwise distance.  

Due to the opaque nature of the ouzo effect, our optimization algorithm \citep{Lee2022} constrains the findings to the time-averaged axisymmetric concentration field. Although the mean field unveils the centerline scaling and the spreading of the profiles, we did not learn much about the temporal dynamics. Moreover, the averaging operation smears out the meandering of the jet, which is vital to identify nucleation near the turbulent/non-turbulent interface (TNTI). TNTI is the interface separating the turbulent region from the non-turbulent one, which has been extensively studied in various kinds of turbulent shear flows, such as in axisymmetric jets \citep{Westerweel2009,Mistry2019}, planar jets \citep{Watanabe2015a,Silva2017}, mixing layers \citep{Attili2014b}, and gravity currents \citep{Krug2015}. Conditioning the flow fields on the distance to the TNTI reveals a sharp gradient of vorticity and concentration across the interface in all the aforementioned studies. \citet{Westerweel2005, Westerweel2009} suggested that the small-scale eddy motion (`nibbling') dominates over inviscid action of the large-scale eddies (`engulfment') as the main entrainment mechanism across the TNTI for an axisymmetric jet. \citet{Watanabe2015a} focused on the TNTI of a reactive planar jet, revealing a significant effect of the interface orientation on the concentration profile. Such an effect is more pronounced for the case with fast reactions. \citet{Lee2022} argued that the measured sub-Gaussian mean concentration profiles result from the meandering TNTI and the accompanying mixing and nucleation.

The opaque nature of the axisymmetric jet of \citet{Lee2022} makes it very difficult, if not impossible, to further investigate the effect of turbulence on the solvent exchange near the TNTI experimentally. We therefore opt for a quasi-two-dimensional jet \citep{Landel2012a, Landel2012b} by confining the jet laterally. Eliminating the depth-integrated information in a 3D jet \citep{Lee2022}, we then can extract information on the turbulent statistics of the scalars. As a reference case, we also injected dyed ethanol to induce quasi-2D turbulent buoyant jets. Utilizing a light attenuation technique, we reconstruct the scalar fields for both dyed ethanol and nucleated oil. We perform conventional and conditional statistics, and compare the spatial profiles and the statistics of the ouzo case and the passive scalar case, revealing the effect of nucleation. Moreover, we develop a very simple model to be able to predict the concentration of the nucleated oil from the passive scalar field. We provide qualitative explanations of the deviation between the model and the measurement.

The paper is structured as follow: In \S2 we detail the experimental setup and the methods to measure the concentration fields. Also, the detection of the TNTI is discussed in \S2. In \S3 we show the results of the conventional and conditional statistics, with supporting discussion and qualitative analysis in \S4. Conclusions and an outlook on future work are presented in \S5.

\section{Experimental method}
\subsection{Set-up}

\begin{figure}[h!tb]
\centering
\includegraphics[scale=1]{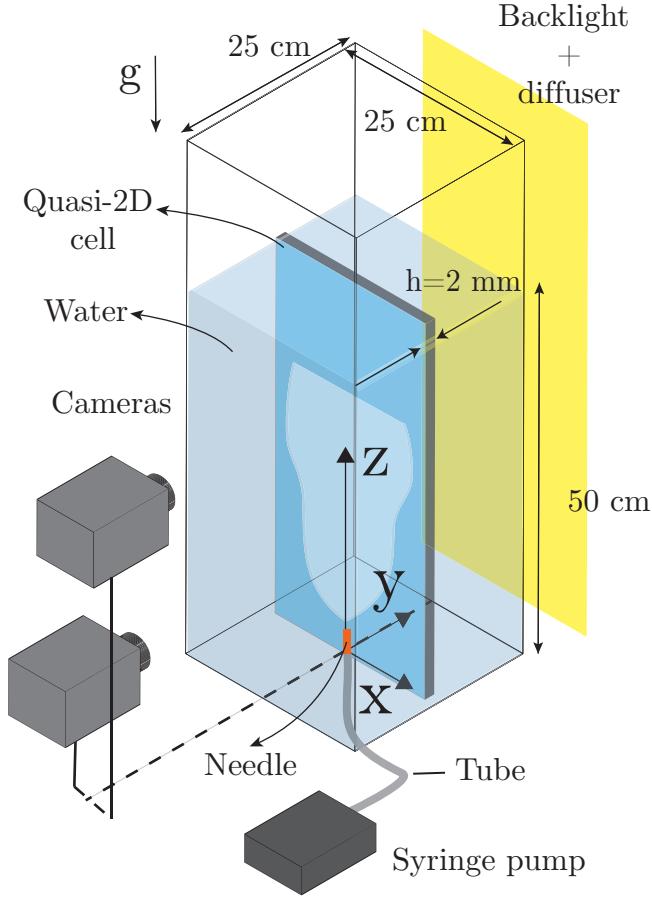} 
\caption{Experimental set-up. The syringe pump injects the ethanol/oil mixture upwards to the quasi-2D geometry filled with water.}
\label{img:setup2}
\end{figure}

We have conducted the experiments in a glass tank with dimensions \SI{25}{\cm} $\times$ \SI{25}{\cm} $\times$ \SI{50}{\cm} (W $\times$ L $\times$ H) as in Fig.\ \ref{img:setup2}. To confine the flow within a quasi-2D geometry, a {{Hele-Shaw}}-like flow cell consisting of two pieces of glass is placed within the tank. The flow cell has a gap of \SI{2}{\mm}, and it is open to the ambient on all four sides to prevent internal circulation. We refill the tank with decalcified water before every run of the experiments. We injected a solution upwards through a round blunt tip needle with an inner diameter d = \SI{0.84}{\mm} and an length l = \SI{25.4}{\mm}. The injected solution consisted of mixtures of ethanol and trans-anethole (Sigma Aldrich, $\geq$99\%) with mass ratio $w_e:w_o=100:1$. As ethanol is the dominant component in the ouzo solution, we also injected dyed ethanol as the reference case, offering a comparison between the common passive scalar jet and the ouzo jet. The fluid was injected by a Harvard 2000 syringe pump at three flow rates, forming quasi-2D turbulent buoyant jets with initial Reynolds numbers ranging from 500 to 800, where buoyancy comes from the approximate 22\% density difference between ethanol and water. The initial Reynolds number is defined as $Re_0=u_0d/\nu$. To minimize the effect of the temperature-dependence of the solubility in the ternary liquids system, the water in the tank and the injected solution were kept at room temperature ($\SI{20}{\celsius}\pm \SI{1}{\kelvin}$) before all the experiments. 

We measure the local concentrations using a light attenuation technique. For the purpose we use a LED backlighting with a diffuser and recorded the degree of light attenuation using two Photron FASTCAM Mini AX200 high-speed cameras with 1024 $\times$ 1024 pixels resolution at 50 fps. The cameras were equipped with Zeiss \SI{100}{\mm} objectives, and installed at two different axial positions to capture the local concentration fields, see e.g. Fig.\ \ref{img:detection}. The cameras were carefully arranged so that the centerline and the TNTI of the jet can both be captured. The injection and the recording lasted for 40--50 s, providing 2000--2500 frames for each experiment. For each condition, the experiment is repeated 3 times for reliable statistical results.

\subsection{Oversaturation}

\begin{figure}
\centering
\includegraphics[scale=1]{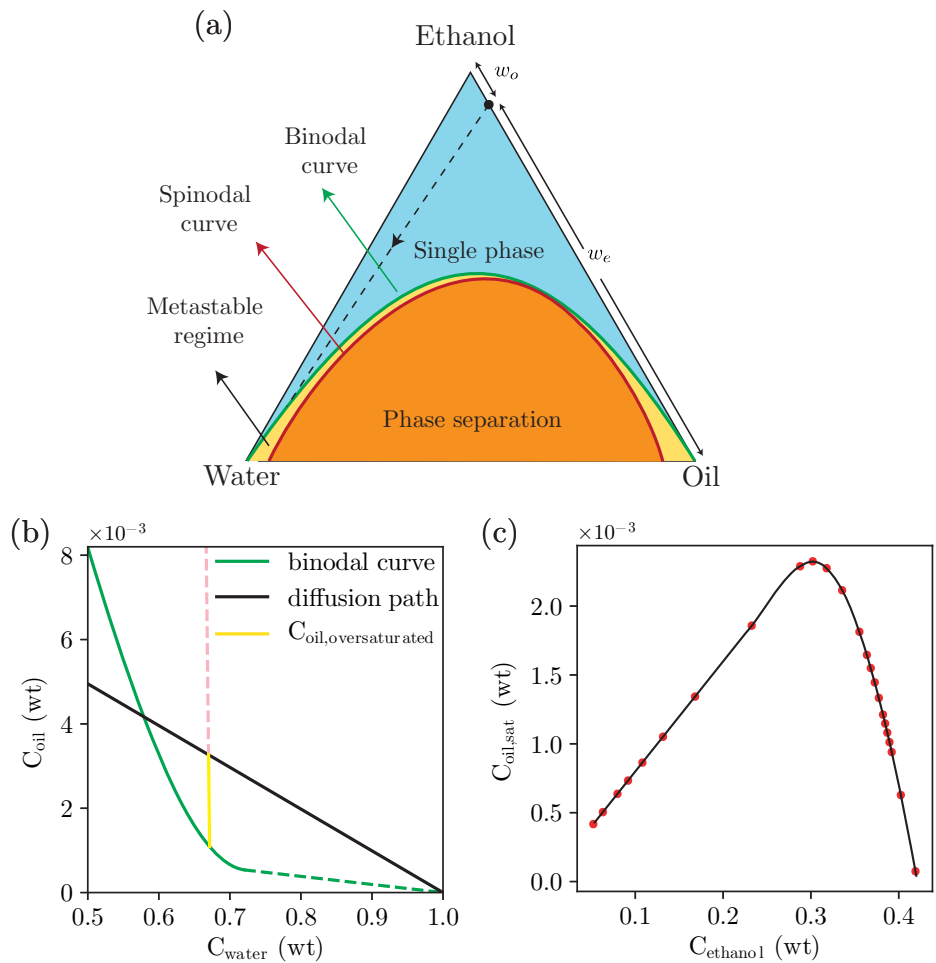}
\caption{Schematic ternary phase diagram and estimation of oversaturation. (a)Ternary phase diagram. The metastable regime in (a) is where the ouzo effect occurs. (b) Binodal curve and the theoretical diffusion path for ouzo mixture ${w_e/w_o = 100}$. The dashed line in (a) corresponds to the black diffusion path in (b). Note that the binodal curve in (b) is determined partly by titration experiments (green solid line) and partly by direct linear approximation (green dashed line). The length of the yellow line segment in (b) measures the oversaturation. The pink dashed line in (b) extends from the yellow line segment to the pure oil phase, which can be approximated with a vertical line considering the tiny amount of oil. (c) Non-monotonic variation of oversaturation as a function of ethanol weight fraction, which is fitted by a $3^{rd}$-order polynomials for $C_{\text{ethanol}}>0.3$, and by a piecewise linear function for $C_{\text{ethanol}}\leq0.3$.} 
\label{img:phasegram}
\end{figure}

To estimate the concentration of the nucleated oil, the oversaturation of the oil needs to be calculated. Fig.\ \ref{img:phasegram}(a) shows the ternary phase diagram for our system. Before the injection, the initial solution lies on the ethanol-oil boundary, close to the ethanol phase. Once mixed with water, the local fluid parcel follows the trajectory indicated by the dashed arrow in Fig.\ \ref{img:phasegram}(a). As the trajectory reaches the green binodal curve, the solution is saturated with oil. Mixing with more water, the local fluid parcel enters the metastable regime and oil droplets start to nucleate. Converting the three-component phase diagram in Fig.\ \ref{img:phasegram}(a) to two components, we obtain Fig.\ \ref{img:phasegram}(b). The green solid line is determined by titration following the procedures detailed in \citet{Lee2022}. The green dashed line represents the part of the binodal curve approximated by a direct linear extrapolation. We do not perform titration here since it is very close to the pure water phase and the titration gets very difficult. The black diffusion path in Fig.\ \ref{img:phasegram}(b) is identical to the black dashed arrow in Fig.\ \ref{img:phasegram}(a), which we assume to be straight. A straight diffusion path suggests that the ethanol/oil ratio remains unchanged during the whole mixing process, which has been discussed in \citet{Li2021} and proved to be a valid assumption. The difference between the diffusion path and the binodal curve, that is, the length of the yellow line segment in \ref{img:phasegram}(b), is the amount of oversaturation. The oversaturation first increases with the water fraction in the local fluid parcel, reaching the peak, and followed by a decline if the water fraction keeps increasing. The decrease in the oversaturation signals the weakening of nucleation, such that dilution overcomes nucleation beyond the optimum mixture composition. Fig.\ \ref{img:phasegram}(c) shows the obtained oversaturation as a function of the local ethanol fraction, where a theoretical maximum oversaturation can be clearly seen around $C_{\text{ethanol}}=0.3$. For $C_{\text{ethanol}}<0.3$ in Fig.\ \ref{img:phasegram}(c), the oversaturation is calculated by the approximated dashed line in Fig.\ \ref{img:phasegram}(b).
\subsection{Calibration}

\begin{figure}[h!tb]
\centering
\includegraphics[scale=1]{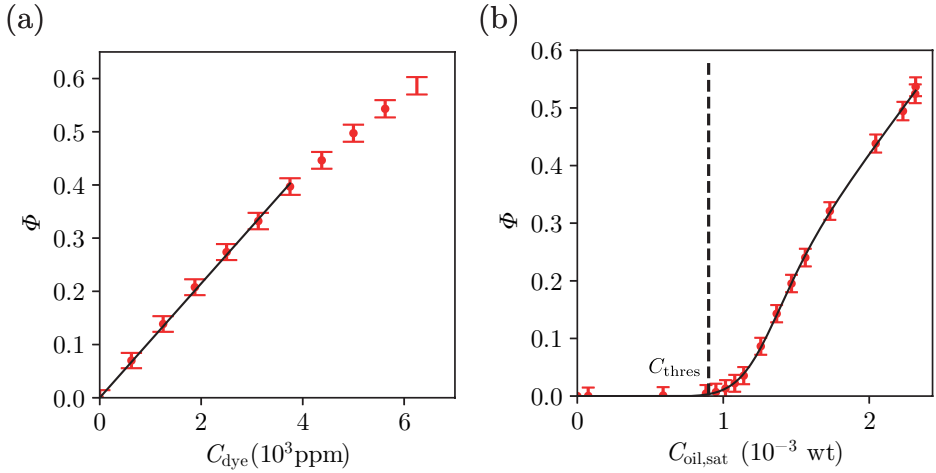}
\caption{Calibration data and the fitted curves. (a) For the reference dye case. (b) For the ouzo case. $\Phi$ is the degree of light attenuation across the 2mm gap, $\Phi = \log(I_\text{{ref}}/I)$. The red data points are obtained by the calibration doses, and the black curves are the fitted curve, which is linear in (a) and an empirical function in fitted by Eq.\ \ref{eq:LA2} in (b). The abscissa in (b) $C_\text{{oil,sat}}$ is the oversaturation of the oil. $C_{\text{thres}}$ denotes the threshold oversaturation used to normalized oil oversaturation.}
\label{img:calicurve2}
\end{figure}

With the definition of the oversaturation, we can apply a light attenuation technique. An in-situ calibration is carried out in a separate closed cell with a \SI{2}{\mm}-gap immersed in the same tank as in Fig.\ \ref{img:setup2}. The cell is first filled with a specific amount of ethanol/oil solution ($w_e:w_o=100:1$), which is transparent and thus serves as the background image. We gradually add doses of water and mix the solution in the cell until it is visually homogeneous, followed by taking 200 frames for each dose. For the reference dye case, the calibration procedure is the same, but the cell is then first filled with pure ethanol followed by adding doses of red food dye. Averaging over 200 frames and dividing by the background image, the average light attenuation level for every pixel is obtained. To reduce the noise, we then average the light attenuation level for every block of 5 $\times$ 5 pixels and take it as one image unit. Relating the logarithmic light attenuation level to the estimated concentration (oversaturation), we obtain the in-situ calibration data for every image unit, see the data points in Fig.\ \ref{img:calicurve2}. 

The red line in Fig.\ \ref{img:calicurve2}(a) shows that a linear calibration curve is established below \SI{4000}ppm, which results in our choice of \SI{3500}ppm as the initial concentration for the dye case. On the other hand, the calibration curve for the ouzo case in Fig.\ \ref{img:calicurve2}(b) seems non-linear and thus we fit it by the empirical function:
\begin{equation}
    \Phi(C) = \log \left( I_{\text{ref}}/I \right)= \frac{a_0}{1+a_1e^{a_2(C-a_3)}}+\frac{a_4(C-a_5)}{1+a_1e^{a_2(C-a_3)}},
\label{eq:LA2}
\end{equation}
where $a_0$, $a_1$, $a_2$, $a_3$, $a_4$, and $a_5$ are the fitting parameters, and $C$ is the oil oversaturation. In Fig.\ \ref{img:calicurve2}(b) we can see that there is a threshold oversaturation, $C_{\text{thres}}$, for the light attenuation to set in. Also, we found in the calibration that the light attenuation level does not monotonically increase with added water, and the peak attenuation level occurs at an oversaturation value, $C_{\text{max}}$, very close to that indicated in Fig.\ \ref{img:phasegram}(c). For a better interpretation of the results, we normalize the oil oversaturation using 
\begin{equation}
\widetilde{C}_{\text{oil}} = \frac{(C_{\text{oil}}-C_{\text{thres}})}{(C_{\text{max}}-C_{\text{thres}})}
\label{eq:cnormoil}
\end{equation}
For the dye case we normalize the concentrations with the initial concentration, written as
\begin{equation}
\widetilde{C}_{\text{dye}} = \frac{C_{\text{dye}}}{C_0}
\label{eq:cnormdye}
\end{equation}

Using the in-situ calibration curves, the concentration fields displayed in Figs. \ref{img:detection}(a,b,d,e) are calculated from the recorded images. 
\subsection{TNTI detection}

\begin{figure}[h!bt]
\centering
\includegraphics[scale=1]{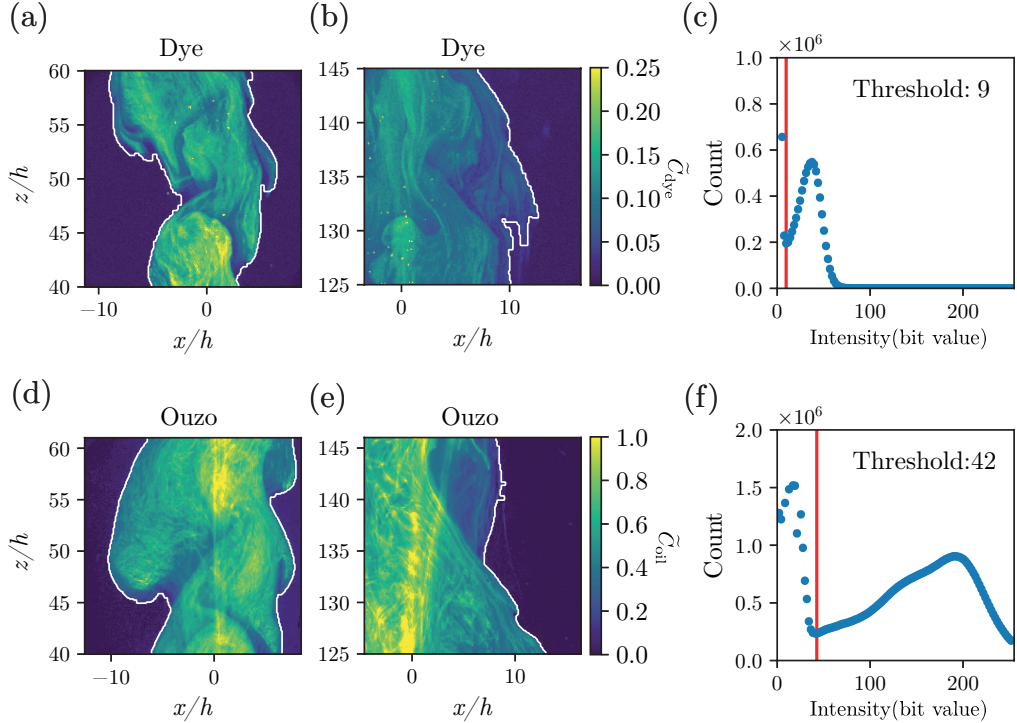}
\caption{Instantaneous scalar fields and TNTI detection for $Re_0=606$. The color code in (a,b,d,e) represents the normalized concentrations. The white contours in (a,b,d,e) indicate the detected TNTI. (a) For the dye case at the low position, $40\leq z/h \leq 60$. (b) For the dye case at the high position, $125\leq z/h \leq 145$. (c) The threshold value for the TNTI detection in case (a). (d) For the ouzo case at the low position, $40\leq z/h \leq 60$. (e) For the ouzo case at the high position, $125\leq z/h \leq 145$. (f) The threshold value for the TNTI detection in case (d).}
\label{img:detection}
\end{figure}

The detection of the TNTI is critical for conditional statistics, and is thus always an essential part of the relevant studies. Following the method presented in \citet{Prasad1989}, we plot the histogram of the bit value across the recorded frames, and identify the first local minimum as the threshold for the TNTI detection, see Figs. \ref{img:detection}(c,f). Figs. \ref{img:detection}(a,b,d,e) show that the detected TNTIs fit reasonably well to the contour of the jets for both the dye and the ouzo case, at both low and height positions. 
For the histogram and the following results on conventional and conditional statistics, we only collect the data when the flow is in a steady state, and we exclude the transients at the start of the injection.  

\section{Results}
\subsection{Conventional statistics}

\begin{figure}[h!bt]
\hspace{-3mm}
\includegraphics[scale=1]{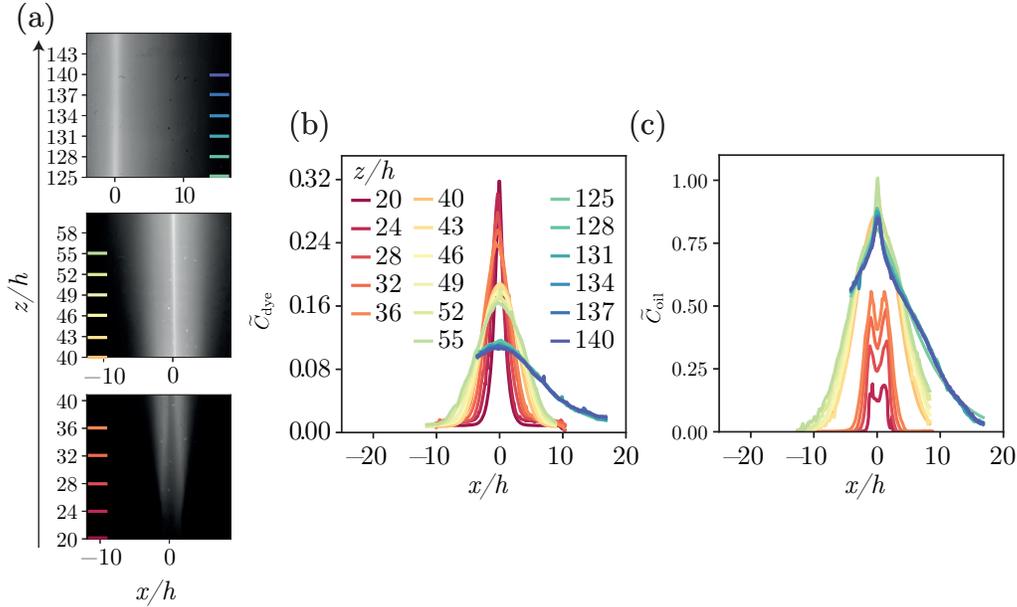}
\caption{Conventional mean concentration profiles for $Re_0=606$. (a) shows the mean concentration fields at three recording views for ouzo case, which serves as a graphical legend for (b,c). (b) is the profiles for dye and (c) for ouzo. Note that the recordings for $20\leq z/h \leq 40$ are not discussed in \S2.4 since these were recorded separately, and we do not perform conditional statistics on these images. The profiles for $125\leq z/h \leq 145$ are cut off on the left-hand side due to the limited recording view, see Figs.\ \ref{img:detection}(b,e).}
\label{img:cmean}
\end{figure}

Fig.\ \ref{img:cmean} displays the streamwise evolution of the normalized mean concentration profiles. Fig.\ \ref{img:cmean}(a) shows a graphical guide for the location of the profiles shown in Figs.\ \ref{img:cmean}(b,c). For the dye case, we see in Fig.\ \ref{img:cmean}(b) evolving Gaussian profiles following continuous mixing and dilution in the streamwise direction. On the other hand, for the ouzo case, the effect of nucleation sets in at $20\leq z/h \leq 40$, exhibiting increasing concentration levels and off-center peaks in Figs.\ \ref{img:cmean}(a,c). The off-center peaks indicate that nucleation is initiated near the TNTI. As the jet travels downstream to $40\leq z/h \leq 60$, the concentrations further increase. Although the off-center peaks are no longer present, we can more or less see off-center humps, suggesting ongoing nucleation events near the TNTI. The concentration profiles from $40\leq z/h \leq 60$ to $125\leq z/h \leq 145$ in Fig.\ \ref{img:cmean}(c) only marginally change, showing the effect of nucleation on reducing dilution. 

\begin{figure}[h!bt]
\centering
\includegraphics[scale=0.9]{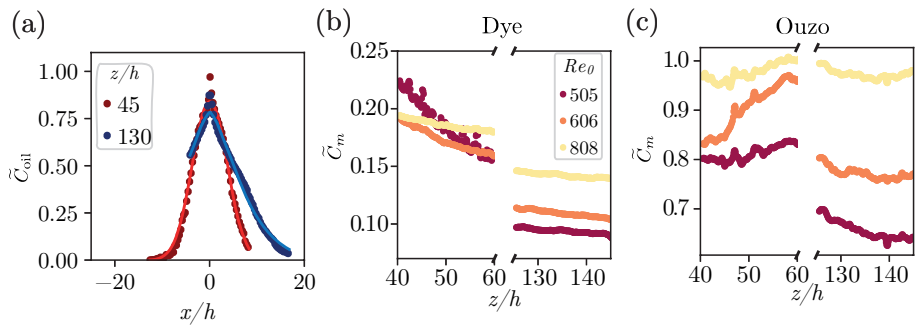}
\caption{Fitted centerline evolution for the concentration profiles in Fig.\ \ref{img:cmean}. (a) demonstrates the deviation between the measured data and the fitted curves near the centerline, which is caused by oil attachment. (b) For the dye cases. (c) For the ouzo cases.}
\label{img:center}
\end{figure}

From the concentration profiles as those in Fig.\ \ref{img:cmean}, we extract the centerline concentrations for all three $Re_0$. However, we think the unproportionally-high centerline concentrations in Figs. \ref{img:cmean}(b) are partly caused by oil attached to the glass wall. We therefore extract the centerline value fitted by the generalized normal distribution function, 
\begin{equation}
\widetilde{C}(x) = \widetilde{C}_m e^{-\frac{|x|^{\kappa}}{\kappa \sigma^{\kappa}}},
\label{eq:fit2}
\end{equation}
where $x$ is the horizontal distance to the centerline, $\widetilde{C}_m$ the centerline concentration, $\sigma$ is the jet width, $\kappa$ a shape parameter, and deviation from $\kappa=2$ indicates non-Gaussianity. The high concentrations near the centerline are excluded when performing the curve-fitting, see the example in Fig.\ \ref{img:center}(a).

Figs. \ref{img:center}(b,c) present the fitted centerline evolution for $40\leq z/h \leq 60$ and $125\leq z/h \leq 145$. Note that the data from the lowest position ($20\leq z/h \leq 40$) is not included as they were recorded separately, and the off-center peaks do not fit the desired generalized normal distribution profile. At first glance, the effect of $Re_0$ in Fig.\ \ref{img:center}(b) is counter-intuitive. For the lowest $Re_0$, the mixing is weakest within $40\leq z/h \leq 60$, and therefore the concentration is the highest for $Re_0=505$. However, what is surprising is that the decay rate of the concentration is fastest for $Re_0=505$ and slowest for $Re_0=808$. We attribute this phenomenon to the limited mixing and the wall friction due to our confining quasi-2D geometry. With the fastest transition toward turbulence and the strongest mixing, the centerline concentration for $Re_0=808$ decays to the lowest value before $z/h = 40$. However, the reduced entrainment due to lateral confinement restricts the mixing and the decay of concentration downstream. This is also why the concentrations within $125\leq z/h \leq 145$ turn out to be highest for $Re_0=808$.

Comparing Fig.\ \ref{img:center}(c) to Fig.\ \ref{img:center}(b), the effect of nucleation can be clearly identified by the increasing concentration at $40\leq z/h \leq 60$. Also, the level of nucleation increases with $Re_0$, as the concentration is highest for $Re_0=808$. The oil droplets nucleate strongest upstream for the highest $Re_0$, and the value almost does not decay from the low to the high position, which results from the restricted dilution and the extended nucleation. 

\begin{figure}[h!tb]
\includegraphics[scale=1]{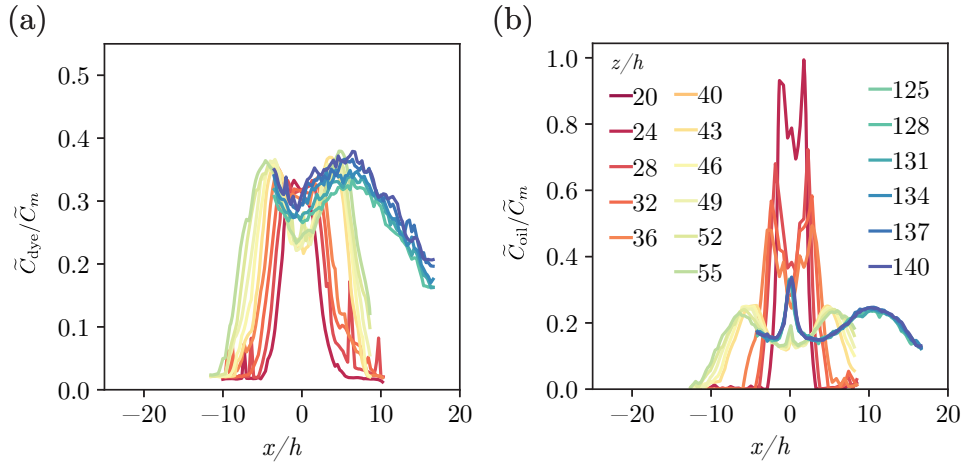}
\caption{Conventional r.m.s. concentration profiles for $Re_0=606$ for (a) dye and (b) ouzo. The profiles for $125\leq z/h \leq 145$ are cut off on the left-hand side due to the limited recording view, see Figs.\ \ref{img:detection}(b,e).}
\label{img:cstd}
\end{figure}

In addition to the mean concentration profiles displayed in Fig.\ \ref{img:cmean}, we show the r.m.s. profiles for $Re_0=606$ in Fig.\ \ref{img:cstd} to complete the analysis on the conventional statistics. The typical off-center peaks for the r.m.s. concentration profiles are present for the dye and the ouzo cases at all three locations. Figs. \ref{img:cstd}(b), however, show that there is a secondary or even the strongest peak near the centerline. We believe that this is also a consequence of the gradual build-up of nucleated oil on the wall, similar to what we see in Figs. \ref{img:cmean}(b).

\begin{figure}[h!tb]
\hspace{-5mm}
\centering
\includegraphics[scale=1]{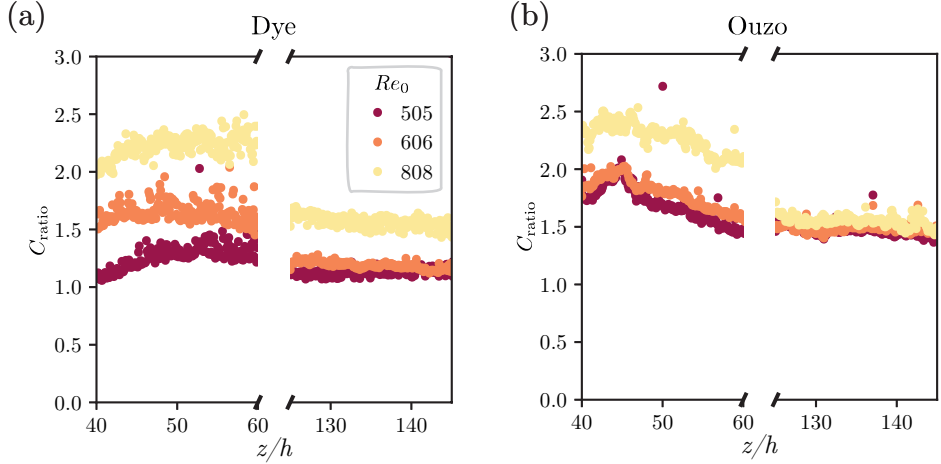}
\caption{Evolution of the ratio of the peak-to-centerline for the r.m.s. profiles in Fig.\ \ref{img:cstd}. (a) For the dye cases, and (b) for the ouzo cases. $C_{\text{ratio}}=\widetilde{C}_{\text{peak}}/\widetilde{C}_{m}$ denotes the ratio between the off-center peak shown in Fig.\ \ref{img:cstd} and the centerline value. Note that for (b) we take the minimum close to the centerline instead of the centerline value itself, as the distinct peak in the centerline shown by the blue curves in Fig.\ \ref{img:cstd}(b) is not a typical feature of r.m.s. profiles, which will lead to an inappropriate comparison.}
\label{img:subpeak}
\end{figure}

To have a deeper look into the off-center peaks featured in Fig.\ \ref{img:cstd}, we extract the ratio between the peak and the centerline concentrations in Fig.\ \ref{img:subpeak}. Comparing Fig.\ \ref{img:subpeak}(b) to Fig.\ \ref{img:subpeak}(a), the concentration ratio for the ouzo case is slightly higher than that for the dye case for $Re_0=505$ and 606, suggesting enhanced fluctuations near the TNTI due to nucleation. As for $Re_0=808$, the most intense nucleation might have occurred before $z/h=40$, causing the off-center peak ratio very close to that as in the dye case.  
\subsection{Conditional statistics}
\begin{figure}[h!tb]
\centering
\hspace{-3mm}
\includegraphics[scale=1]{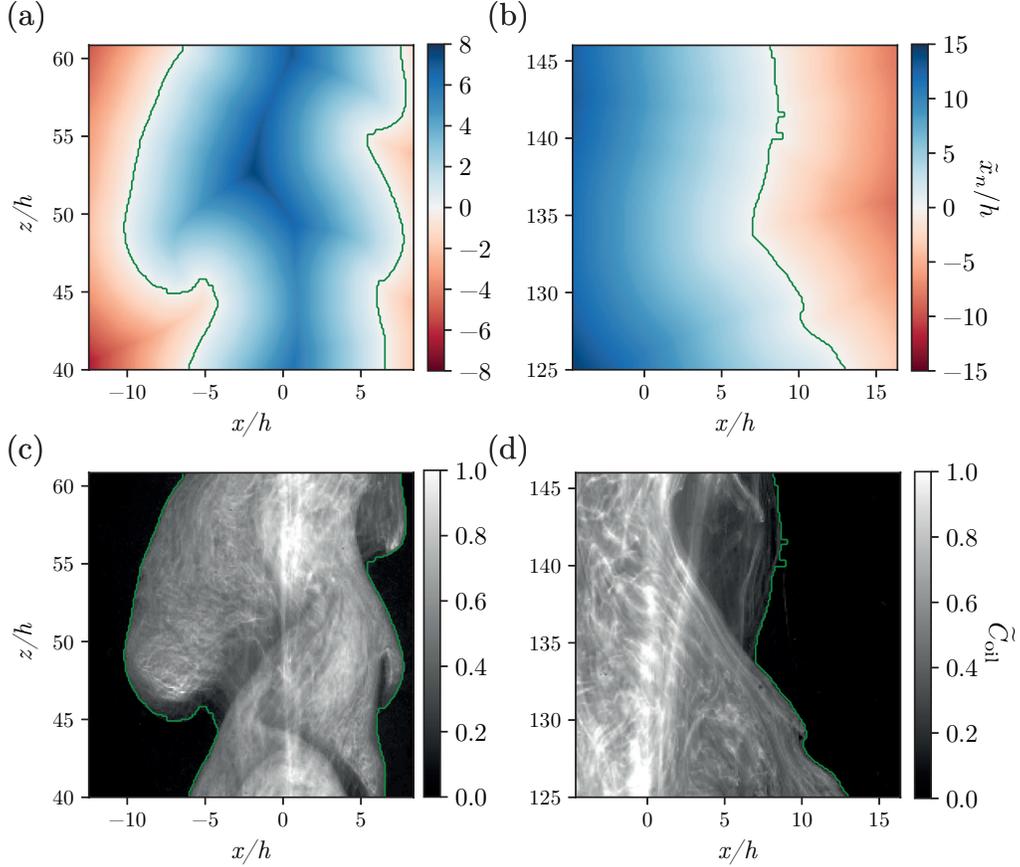}
\caption{Conditional distance to the TNTI for the ouzo case with $Re_0=606$. (a) $40\leq z/h \leq 60$ and (b) $125\leq z/h \leq 145$. (c,d) The concentration fields corresponding to (a,b), respectively. The green contours mark the TNTI.}
\label{img:dist}
\end{figure}

While the conventional statistics have already provided us with insights into the solvent exchange process of a turbulent jet, the information near the TNTI is smeared due to temporal averaging. Using the detected TNTI in Fig.\ \ref{img:detection}, we utilize a distance transform on the TNTI for the conditional statistics, see the examples in Fig.\ \ref{img:dist}. We define $\tilde{x}_n$ as the distance to the TNTI, with $\tilde{x}_n>0$ in the turbulent region and $\tilde{x}_n<0$ in the non-turbulent region.

Since the conditional profiles look similar for all three $Re_0$, in the following analysis, we only show the results for $Re_0=606$. As for the biased centerline concentrations due to nucleation on the wall we discussed in \S.3.1, we exclude those data points with the help of a joint histogram of the concentrations and the scalar dissipation rate. The troublesome data are almost separate from most of the data points on the 2D histogram, and thus we can cut them off as they are clear outliers.  

\begin{figure}[h!tb]
\centering
\includegraphics[scale=1]{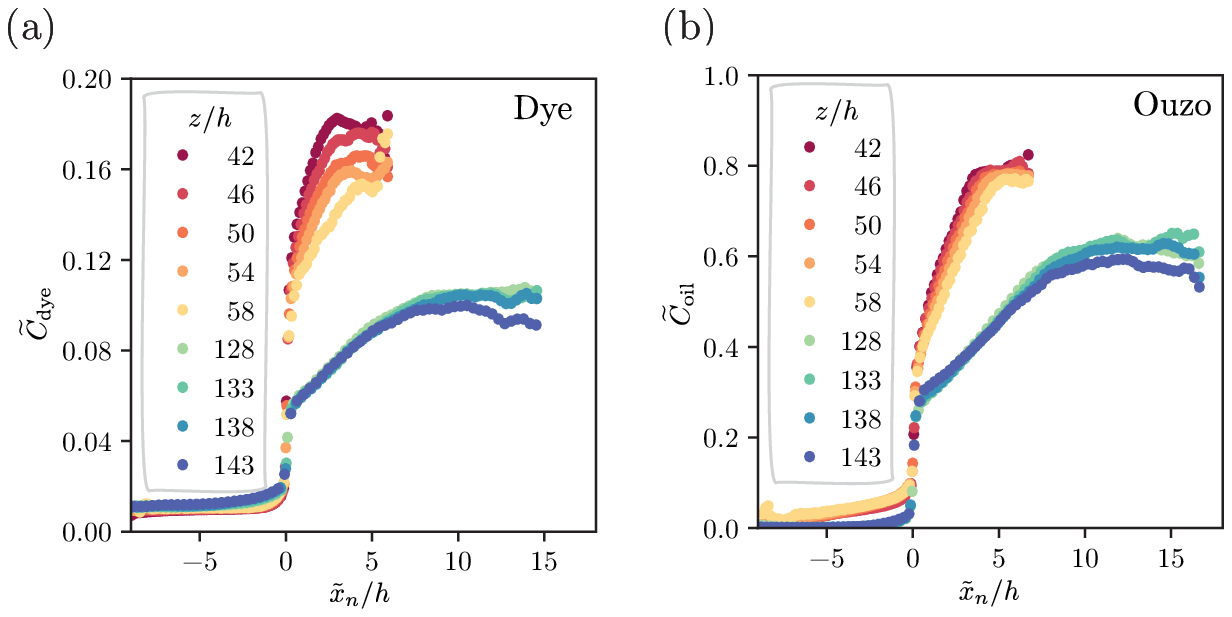}
\caption{Mean concentration profiles conditioned on the distance to the TNTI for $Re_0=606$. (a) dye and (b) ouzo.}
\label{img:meanprof}
\end{figure}

For the conditional mean concentration profiles, the existence of a sharp concentration gradient across the TNTI has been widely confirmed. The profiles in Fig.\ \ref{img:meanprof} show that we successfully capture this feature for both the dye case and the ouzo case. Similar to what we observe in the conventional mean profiles in Fig.\ \ref{img:cmean}, nucleation of the oil serves as a source term, competing with the dilution. Fig.\ \ref{img:meanprof}(b) illustrates not only the almost unchanged concentration levels for low positions, but also the reduced dilution from low to high positions, as compared to the dye case in Fig.\ \ref{img:meanprof}(a). Focusing on the turbulent region, $\tilde{x}_n>0$, most of the profiles in Fig.\ \ref{img:meanprof} extend beyond the peak concentration, reaching a plateau or starting to decline. This is caused by the frames with part of the TNTI out of the scope, which leads to the large $\tilde{x}_n>$ value marked by the dark blue color in Fig.\ \ref{img:dist}(b). 

\begin{figure}[h!tb]
\centering
\includegraphics[scale=1]{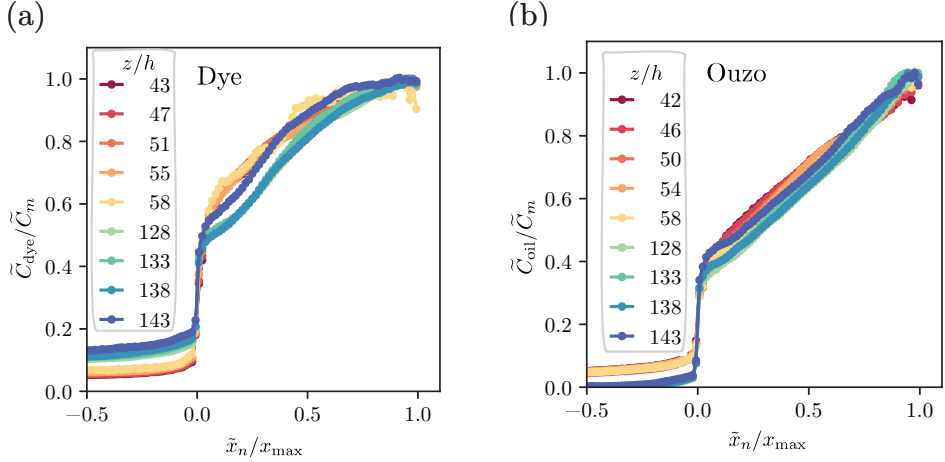}
\caption{Normalized mean concentration profiles conditioned on the distance to TNTI for $Re_0=606$. (a) dye and (b) ouzo. The concentrations are normalized by the centerline value $\widetilde{C}_{m}$, and the distance is normalized by the distance from the centerline to the TNTI $x_{\text{max}}$.}
\label{img:normmeanprof}
\end{figure}

To facilitate a fair comparison for the dye and the ouzo case, we further normalize the mean profiles by the corresponding centerline concentrations, $\widetilde{C}_m$, and the distance from the TNTI to the centerline, $x_{\text{max}}$, see the axes of Fig.\ \ref{img:normmeanprof}. Note that the normalization with $x_{\text{max}}$ automatically excludes the beyond-the-peak plateau mentioned in the previous paragraph. Also, the centerline concentrations used here originate from the fitting results presented in Fig.\ \ref{img:center}. Fig.\ \ref{img:normmeanprof} shows that the profiles collapse well across the height for $40\leq z/h \leq 145$ for both the dye and the ouzo case. Focusing on the gradient across the TNTI, we intuitively expected to see a larger gradient for the ouzo case due to nucleation. Surprisingly, Fig.\ \ref{img:normmeanprof} shows that the magnitude for the ouzo case is smaller than that for the dye case, which results from the higher centerline concentrations we use to normalize the profiles. 

Fig.\ \ref{img:normmeanprof}(b) also shows that the concentration gradient for the ouzo case is larger than that for the dye case within the turbulent region ($\tilde{x}_n>0$). Together with what we just discussed near the TNTI, it is revealed that the nucleation events are not restricted to the region near the TNTI. Instead, it gets stronger as we move towards the centerline within the turbulent region. 

\begin{figure}[h!tb]
\centering
\includegraphics[scale=1]{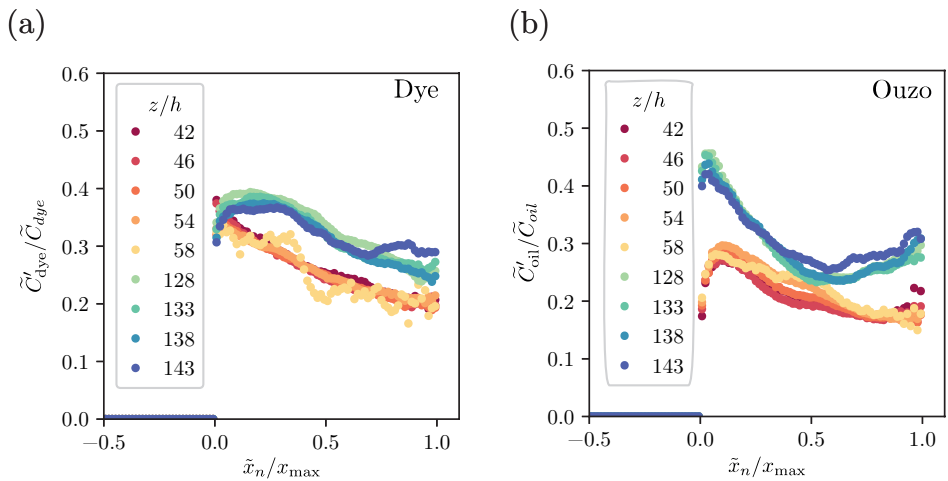}
\caption{Normalized concentration fluctuation profiles conditioned on the distance to TNTI for $Re_0=606$. (a) dye and (b) ouzo. The concentration r.m.s. are normalized by the mean values shown in Fig.\ \ref{img:meanprof}, and the distance is normalized by the distance from the centerline to the TNTI. Note that we force the value in the non-turbulent region ($\tilde{x}_n<0$) to be zero as the fluctuation there is only noise.}
\label{img:stdprof}
\end{figure}

While the mean concentration profile exhibits the outcome of the solvent exchange process upon mixing, the fluctuations of the concentration in time and space are also crucial to understand the process. Fig.\ \ref{img:stdprof} presents the r.m.s. profiles normalized by the mean profiles in Fig.\ \ref{img:meanprof}, that is, the relative fluctuation of concentrations in time. For the dye case in Fig.\ \ref{img:stdprof}(a), the fluctuation decreases monotonically from the TNTI to the centerline, resulting from the relatively intense mixing across the TNTI. On the other hand, the non-monotonic variation at the low positions in Fig.\ \ref{img:stdprof}(b) reveals that the peak fluctuations are shifted slightly toward the turbulent region, where nucleation might be the strongest. The blue curves in Fig.\ \ref{img:stdprof}(b) also show that the relative fluctuations increase significantly as the jet travels downstream, especially near the TNTI and the centerline. We attribute the enhanced fluctuations near the centerline and near the TNTI to the continuous nucleation of the oil droplets.

\begin{figure}[h!tb]
\centering
\includegraphics[scale=1]{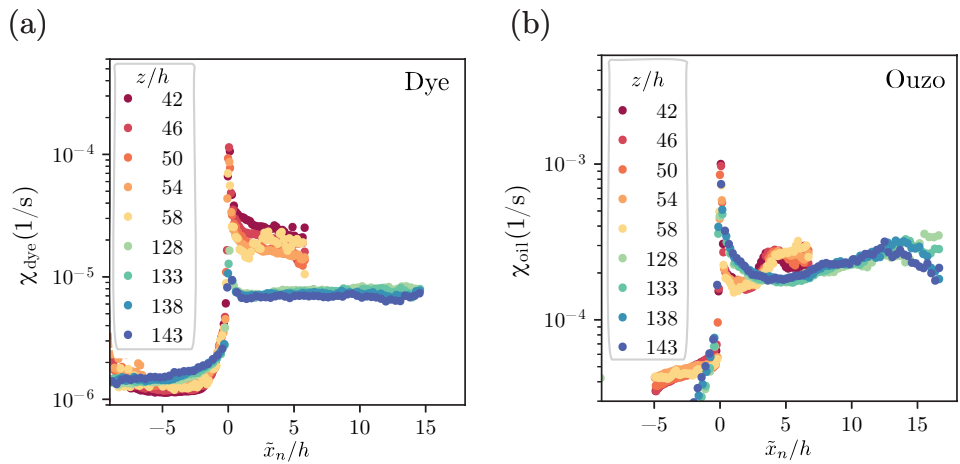}
\caption{Mean scalar dissipation rate profiles conditioned on the distance to TNTI for $Re_0=606$. (a) dye and (b) ouzo.}
\label{img:dispprof}
\end{figure}

To study the spatial fluctuation of the concentration fields, we calculate the conditional (on distance from the TNTI) mean scalar dissipation rate

\begin{equation}
\chi = 2D\sum\limits_{i=1,2}\left(\frac{\partial C}{\partial x_i}\right)^2
\label{eq:disp}
\end{equation}
where $C$ is the local concentration, $D$ is the diffusion coefficient of ethanol, and $x_1$ and $x_2$ denote the $x$ and $z$ directions in the recorded view. Fig.\ \ref{img:dispprof} shows that we capture the peak dissipation rate near the TNTI, which has been reported in the literature \citep{Attili2014b}. As the jet travels downstream and mixes with the ambient water, the scalar dissipation rate naturally decreases, see the decay from the red to the blue curves in Fig.\ \ref{img:dispprof}(a). However, for the ouzo case, $\chi$ remains almost unchanged from the low to the high position, suggesting that nucleation sustains the spatial gradient of the concentrations in the entire span of the jet, especially on the two ends, namely the TNTI and the centerline. This finding further supports our argument that mixing and nucleation occur within the entire turbulent region and last till a very downstream position.

\begin{figure}[h!tb]
\centering
\includegraphics[scale=1]{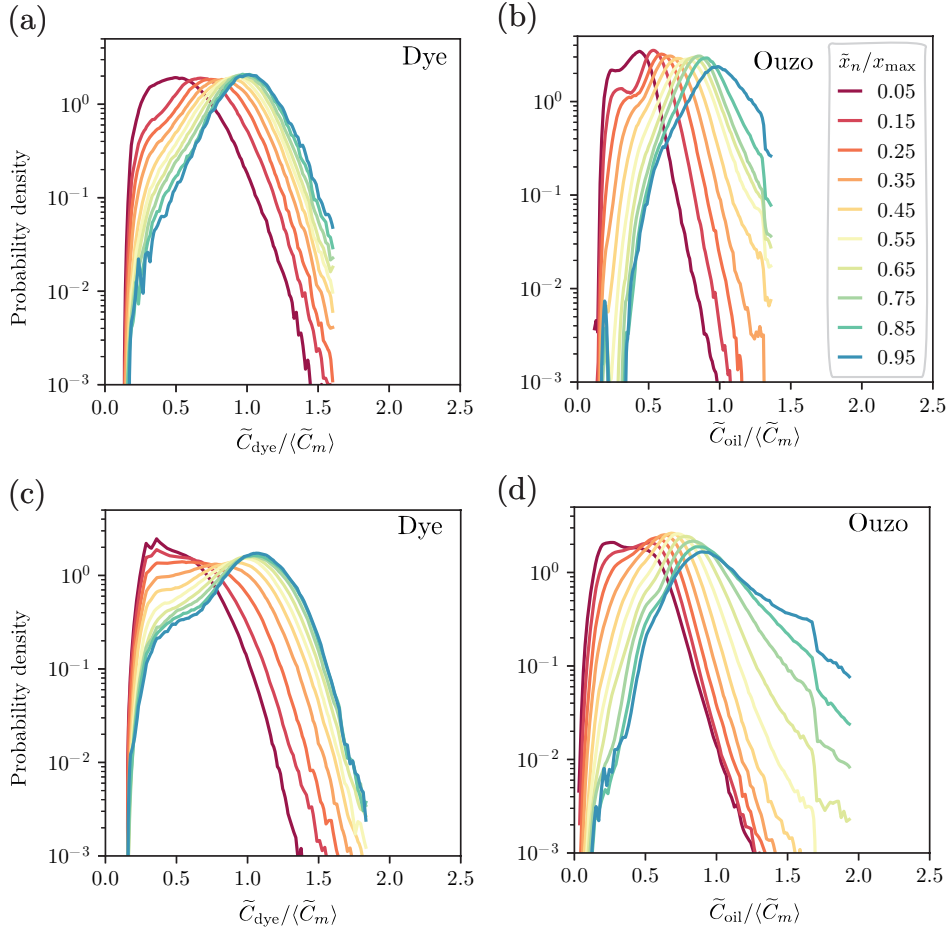}
\caption{PDFs for the concentrations conditioned on the distance to TNTI for $Re_0=606$. The color code given in (b) holds for all four plots. (a,c) dye and (b,d) ouzo. (a,b) at $40\leq z/h \leq 60$. (c,d) at $125\leq z/h \leq 145$. $\tilde{x}_n/x_{\text{max}}$ denotes the distance to the TNTI normalized by the local jet width. The value increases from 0 near the TNTI to 1 near the centerline. The concentrations are normalized by the mean centerline values in the entire domain (20h $\times$ 20h), $\langle \widetilde{C}_m \rangle $.}
\label{img:pdf}
\end{figure}

To evaluate the interaction between turbulent mixing and nucleation, the probability density functions (PDFs) of the concentration field can provide further insights. In Fig.\ \ref{img:pdf} we present the histogram of the conditional concentrations for the dye and the ouzo cases, at low and high positions. While the data is conditioned on the distance to the TNTI for each case, the height dependence is ignored within the domain. Fig.\ \ref{img:pdf}(a) shows bimodal distribution profiles close to the TNTI (red and orange curves) for the dye case at the low position, which indicates that the entrained water is not fully mixed. On the other hand, for the same position in the ouzo case, the red curves in Fig.\ \ref{img:pdf}(b) exhibit more pronounced bimodal profiles close to the TNTI. We attribute the shape to nucleation in the diluted area near the TNTI, which enhances the bimodal curves for the dye case. The left peak of the bimodal shape for the red curves in Fig.\ \ref{img:pdf}(b) shows the presence of the unmixed water, consistent with the red curves in Fig.\ \ref{img:pdf}(a). The local minimum and the right peak indicate the effect of nucleation on the concentration distributions. 

Moving to the high position, Figs. \ref{img:pdf}(c, d) show that nucleation positively skews the distribution across the jet, especially close to the centerline. One more noticeable thing is that the spread between the profiles for the ouzo case in Figs. \ref{img:pdf}(b,d) is significantly larger than the dye case in Figs. \ref{img:pdf}(a,c). Such a larger spread again proves that the nucleation events penetrate deeply into the centerline, aligned with the large concentration gradient within the turbulent region shown in Fig.\ \ref{img:normmeanprof}(b).

Up to this point, we have been discussing the effect of the solvent exchange by comparing the concentration field of the ouzo case to the passive scalar case. However, we would like to emphasize that the concentration field of the ouzo case is in fact linked to the passive scalar case (ethanol concentrations). In other words, the turbulent mixing between the (dyed) ethanol in the jet fluid and the ambient water leads to the nucleation of the oil droplets for the ouzo case. To relate the nucleated oil concentration field to the passive scalar one, we condition the scalar dissipation rate on the local concentrations. \citet{Mellado2009} and \citet{Sutton2013} presented a non-monotonic variation and a local optimum of the dissipation rate in the mixture fraction parameter space, where the mixture fraction measures the mass fraction of the fuel stream in the fuel/air mixture. As the optimum can be interpreted as the concentration for a fast reaction rate \citep{Bilger2004}, we argue that we can replace the reaction rate with the nucleation rate in our study, at least qualitatively, as the ouzo effect is generally considered as a spontaneous emulsification process \citep{Ruschak1972, Sitnikova2005}. 

\begin{figure}[h!tb]
\centering
\includegraphics[scale=1]{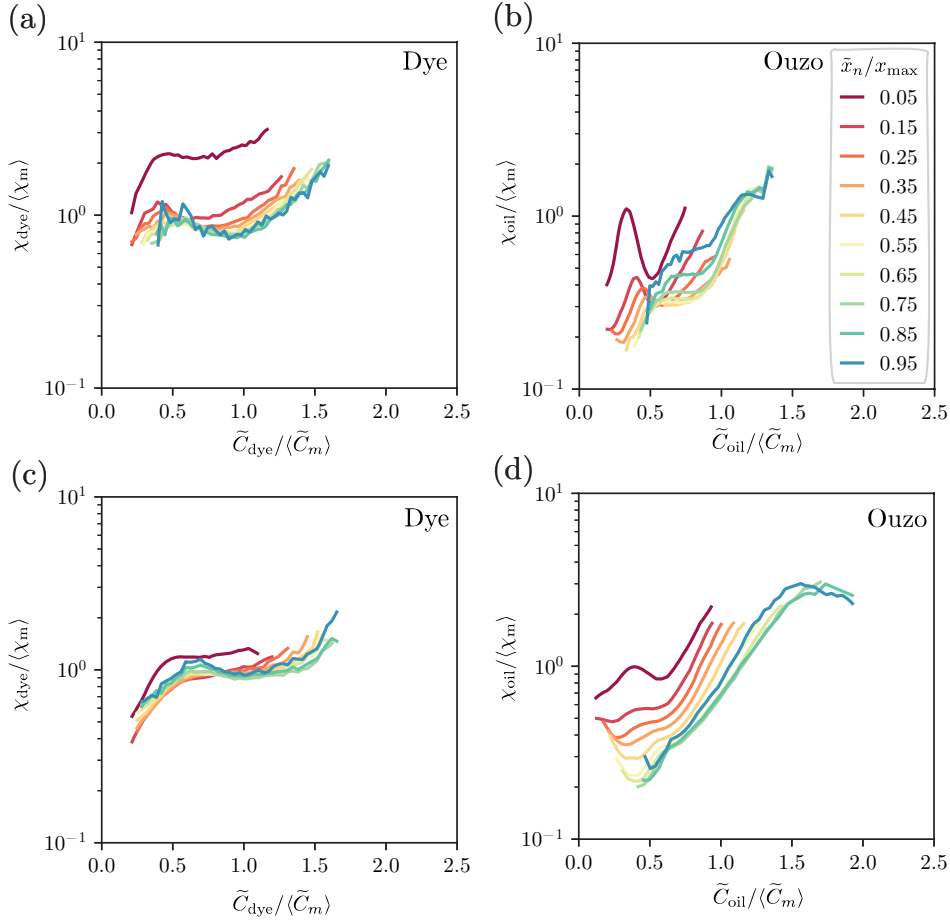}
\caption{Scalar dissipation rates conditioned on the concentration and the distance to TNTI for $Re_0=606$. (a,c) dye and (b,d) ouzo. (a,b) at $40\leq z/h \leq 60$. (c,d) at $125\leq z/h \leq 145$. The concentrations are normalized the same way as that in Fig.\ \ref{img:pdf}, and the dissipation rates are normalized by the mean value in the entire domain (20h x 20h), $\langle \chi_m \rangle$.}
\label{img:dispc}
\end{figure}

In Fig.\ \ref{img:dispc}(a) we identify a local maximum of $\chi$ at low concentration, which indicates intense mixing between the jet fluid and the ambient water for the passive scalar. The scalar dissipation rate of the passive scalar eventually creates the concentration field for the ouzo case, leading to all the results for the ouzo case in Fig.\ \ref{img:meanprof} -- Fig.\ \ref{img:pdf}. The solvent exchange process then in turn builds up the spatial gradient, delivering the results in Fig.\ \ref{img:dispc}(b). The pronounced local maximum near the TNTI in Fig.\ \ref{img:dispc}(b) indicates nucleation of a modest amount of the oil following entrainment and mixing, corresponding to the local minimum in Fig.\ \ref{img:pdf}(b). At the downstream position, Fig.\ \ref{img:dispc}(c) shows smeared spatial gradient with sufficient mixing, with almost collapsing curves across the jet. Fig.\ \ref{img:dispc}(d), on the other hand, exhibits a large spread of the scalar dissipation rate across the jet. Approaching the centerline, the dissipation rate increases almost monotonically with the concentration, proving the continuous nucleation at the downstream position.   

\section{Discussion}
\subsection{Scalar flux and nucleation rate near TNTI}
In \S3.2 and Fig.\ \ref{img:normmeanprof}, we present a qualitative description of the conditional mean concentration profiles. To gain quantitative insights from the concentration gradient, we apply a theoretical framework adapted from \citet{Westerweel2009}. \citet{Westerweel2009} estimated the momentum flux and the scalar flux transported across the TNTI using the conditional profiles, see Figs. 2 \& 13 and \S2.1 \& \S4.5 in their paper for the details. Using a control volume analysis, the momentum and scalar transport can be written as
\begin{align}
& E_b\Delta U \approx \Delta \tilde{\tau} \approx -\nu_e\frac{U_s}{\sigma_x}, \\
& E_b\Delta C \approx \Delta \tilde{F} \approx -D_e\frac{C_s}{\sigma_x},
\label{eq:cv}
\end{align}
where $E_b$ is the velocity of the interface toward the non-turbulent region, $\Delta U$ and $\Delta C$ denotes the jump or the sharp velocity and concentration difference across the TNTI, $\tilde{\tau}$ and $\tilde{F}$ is the momentum flux (stress) and the scalar flux, $\nu_e$ and $D_e$ is the eddy viscosity and the eddy diffusivity, $U_s$ and $C_s$ is the velocity and concentration difference within the turbulent region, and $\sigma_x$ is the width of the turbulent region.

Using Eq.\ \ref{eq:cv} and the approximated relation between the eddy viscosity and eddy diffusivity derived from rapid distortion theory, they suggested
\begin{equation}
\left(\frac{\Delta U}{U_s}\right)\left(\frac{\Delta C}{C_s}\right)^{-1} \approx \frac{\nu_e}{D_e} \approx 0.5
\label{eq:eddydiff}
\end{equation}

Substituting Eq.\ \ref{eq:eddydiff} into Eq.\ \ref{eq:cv}, we get
\begin{equation}
\frac{\Delta U}{U_s} = \frac{\Delta C}{2C_s} = \frac{-D_e}{2E_b\sigma_x}
\label{eq:fluxlink}
\end{equation}

For the dye case, Eq .\ref{eq:fluxlink} can be written as 
\begin{equation}
\frac{-D_e}{E_b} = \sigma_x\frac{\Delta C_{dye}}{C_{s,dye}}
\label{eq:dyeflux}
\end{equation}

For the ouzo case, to estimate the concentration flux across the TNTI due to nucleation, we introduce a source term $\dot{P}$ into Eq.\ \ref{eq:cv}
\begin{equation}
E_b\Delta C  = -\left(D_e\frac{C_s}{\sigma_x}+\dot{P}\right)
\label{eq:ouzoflux}
\end{equation}

Substituting $D_e$ using Eq.\ \ref{eq:dyeflux}
, we get
\begin{equation}
\dot{P} = -E_b\left(\Delta C_{oil}+ C_{s,oil}\frac{\Delta C_{dye}\sigma_{x,dye}}{C_{s,dye}\sigma_{x,oil}}\right)
\label{eq:nucrate}
\end{equation}

\begin{figure}[h!tb]
\centering
\includegraphics[scale=1]{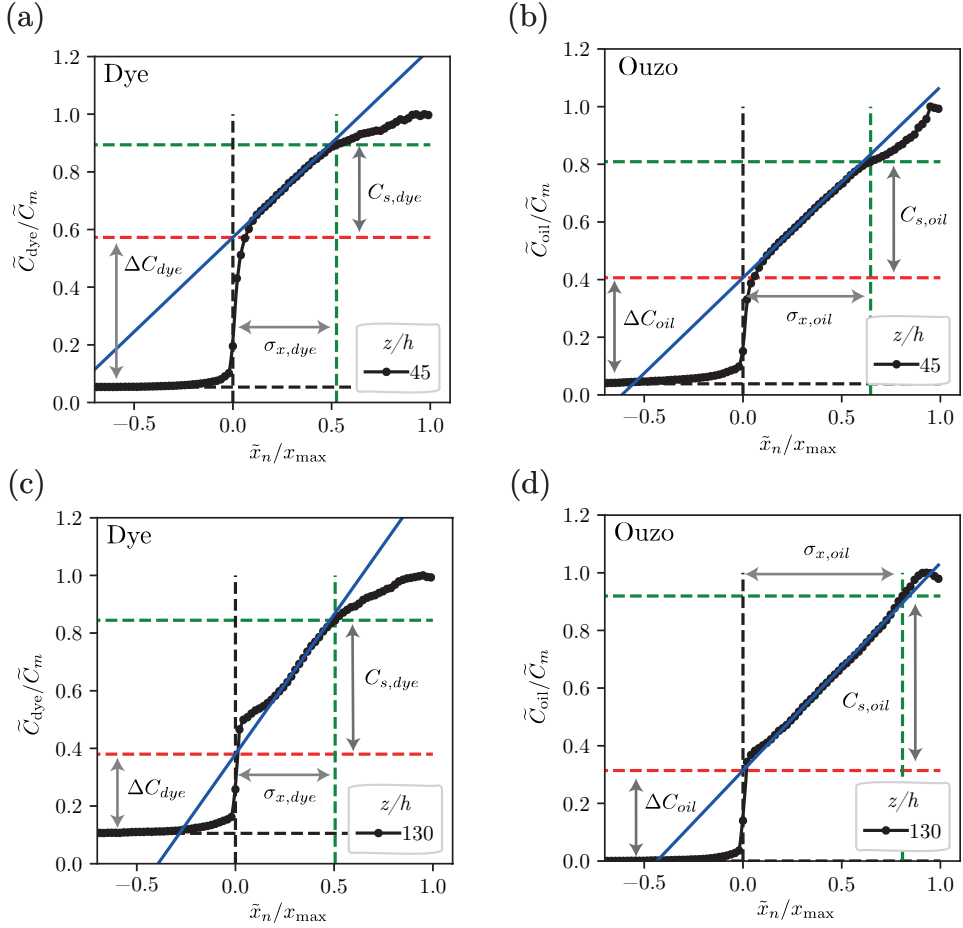}
\caption{Normalized concentration profiles conditioned on the distance to TNTI for $Re_0=606$. The green and red dashed lines mark the range where we estimate the scalar flux, and the blue lines indicate the associated concentration gradient. The (a,c) dye and (b,d) ouzo. (a,b) At $z/h = 45$. (c,d) At $z/h = 130$.}
\label{img:jump}
\end{figure}

\begin{figure}
\centering
\includegraphics[scale=0.8]{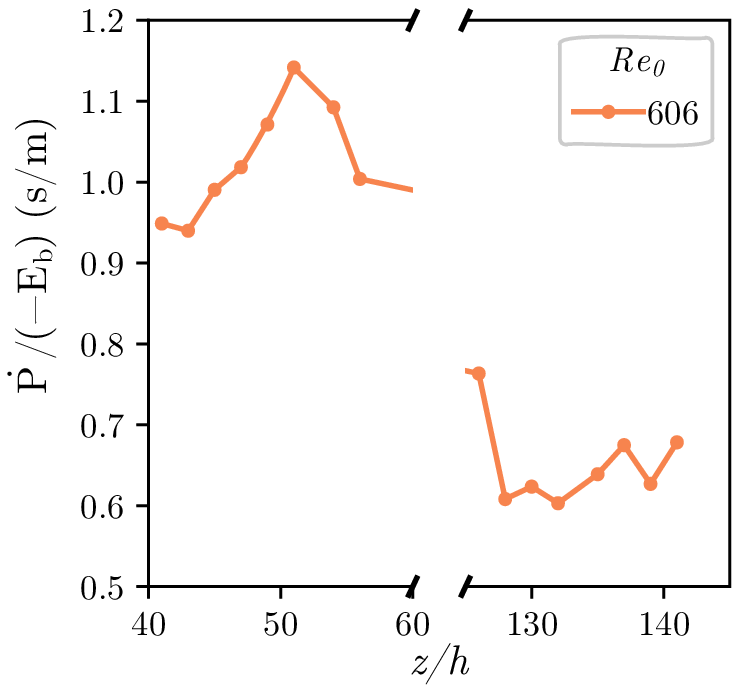}
\caption{Estimated nucleation rate divided by the propagating velocity of the TNTI for $Re_0=606$.}
\label{img:nucrate}
\end{figure}

All the parameters in Eq.\ \ref{eq:nucrate} can be obtained by the normalized conditional profiles shown in Fig.\ \ref{img:jump}, except for $E_b$. The source term divided by $E_b$ can then be calculated at different heights, $z/h$, delivering the estimated nucleation rate divided by $-E_b$ shown in Fig.\ \ref{img:nucrate}. Although we are not able to accurately measure $E_b$, as the time-averaged jet width doesn't increase at the high position, it is fair to estimate that $-E_b$ decreases with height. Therefore, it is certain that the nucleation rate $\dot{P}$ at the low ($40\leq z/h \leq 60$) is larger than that at the high position ($125\leq z/h \leq 145$), which indicates that nucleation near the TNTI is significantly more intense upstream. The analysis here provides quantitative information about the nucleation rate near the TNTI.   

\subsection{Modelling concentration of the nucleated oil}
\begin{figure}
\centering
\includegraphics[scale=1]{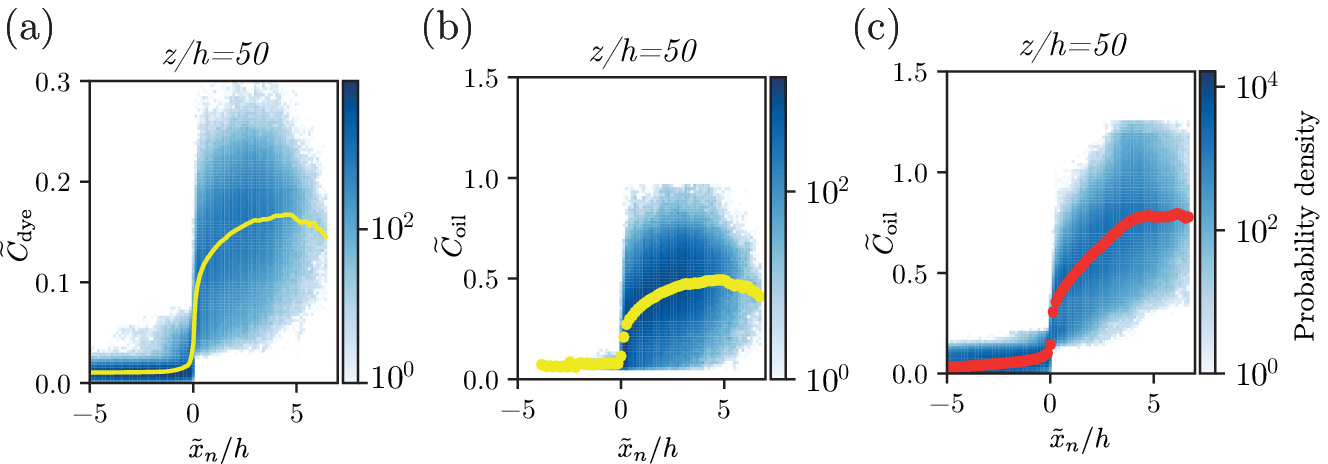}
\caption{Joint histogram of the concentration distribution and the mean concentration profiles conditioned on the distance to TNTI for $Re_0=606$ at $z/h=50$. The yellow and red profiles are the averaged values of the corresponding joint histogram. (a) For dye. (b) Oversaturated oil scalar field modelled from (a) and the phase diagram in Fig.\ \ref{img:phasegram}(b). (c) Measured oversaturated oil scalar field.}
\label{img:profconv}
\end{figure}

As mentioned in \S3.2, we believe that the concentration field of the ouzo case is linked to the concentration field of the passive scalar case, which is in fact the ethanol concentration. Therefore, we take one step further, estimating the concentration field of the nucleated oil using the passive scalar field and the red curve in the phase diagram presented in Fig.\ \ref{img:phasegram}(c). This modelling approach converts the ethanol concentration histogram in Fig.\ \ref{img:profconv}(a) to the oil concentration histogram in Fig.\ \ref{img:profconv}(b). Compared to the measured result in Fig.\ \ref{img:profconv}(c), the modelled oil concentration in Fig.\ \ref{img:profconv}(b) is obviously lower. 

We attempt to identify the possible causes for such a deviation. First, the converting curve in Fig.\ \ref{img:phasegram}(c) cannot capture the threshold oversaturation to induce nucleation, see the discussion for the calibration curve in Fig.\ \ref{img:calicurve2}(b). As a result, the sharp concentration gradient across the TNTI significantly reduces in Fig.\ \ref{img:profconv}(b). 

Second, $\widetilde{C}_{oil}$, the rescaled oversaturation from the measurements exceeds the theoretical maximum, 1, which is certainly not predicted by the model. The over-estimated oversaturation reveals the limitation of the light attenuation technique, which was also observed in our previous work on the round jet \citep{Lee2022}. We attribute such a limitation to the effect of droplet size on the light attenuation level, which is neglected in the measurement. With the in-situ calibration approach, we automatically assume the droplet size is monodisperse and remains the same in the calibration and the turbulent jet. While we can't resolve the oil droplets, we argue that the oil droplets nucleated in the jet might be smaller than those in the calibration, especially in the region with intense mixing and nucleation. The size inconsistency compromises the accuracy of the measurement. Bearing this in mind, the oversaturation we measure here is essentially an apparent concentration of the nucleated oil. The apparent concentration includes the effect of the droplet size, which requires future investigation to disentangle.

Last but not least, the modelling framework relies on a phenomenological assumption. The nucleated oil droplets must stay in the same local fluid parcel, such that the phase diagram and the ethanol concentrations can be exploited to calculate the local oil concentrations. As long as the oil droplets separate from the local fluid parcel, the oil concentration can no longer be estimated by the ethanol concentration, which compromises the phenomenological assumption.  

\section{Conclusions}
We have experimentally investigated the concentration field induced by the solvent exchange in a quasi-2D turbulent buoyant jet. We adapt the light attenuation technique applied in \citet{Lee2022} to estimate the local concentration of the nucleated oil droplets. The quasi-2D geometry not only simplifies the data processing algorithm in \citet{Lee2022}, but also overcomes the limitation of the opaque flow. The geometry enables us to obtain all the instantaneous concentrations and thus the turbulent statistics, which is a step further from the mean field in our previous work \citep{Lee2022}. We have further obtained the concentration profiles and the statistics conditioned on the distance to the TNTI, where the turbulent entrainment and mixing are initiated \citep{Westerweel2009,Mistry2019}. The conditional profiles provide physical insights about the spatial distribution of the solvent exchange process in the turbulent jet flows.

In the streamwise direction, the increasing concentrations upstream and the reduced dilution downstream clearly reveal the effect of nucleation. Using a control volume analysis, we quantitatively show that the nucleation rate near the TNTI is higher at the low position, and decreases as the jet travels downstream to the high position. However, the nucleation events are not restricted to the TNTI. Instead, they penetrate deeply into the turbulent region, evidenced by the steep conditional mean profiles in Fig.\ \ref{img:normmeanprof}(b), the enhanced fluctuation in Fig.\ \ref{img:stdprof}(b), the sustained dissipation rate in Fig.\ \ref{img:dispprof}(b), and the wide-spread PDFs in Figs. \ref{img:pdf}(b,d). 

Although the modelling attempt does not deliver satisfying results, we identify the possible causes for the deviation between the measurements and the model, including the limitations of the measurement technique and the phenomenological assumptions. While the concentration of the nucleated oil is closely connected to the passive scalar field (ethanol concentrations), the spatial and temporal trajectories of the nucleated droplets play a vital role in determining their concentrations.

This experimental investigation unveils the complexity of the solvent exchange in a quasi-2D turbulent jet. Despite being mostly qualitative, the results still serve as stepping stones for the future modelling efforts and the droplet-resolved experimental improvement. 

\section*{Acknowledgement}
The authors acknowledge the funding by ERC Advanced Grant Diffusive Droplet Dynamics (DDD) with Project No.740479 and Netherlands Organisation for Scientific Research (NWO) through the Multiscale Catalytic Energy Conversion (MCEC) research center.

We thank G.W. Bruggert, M. Bos, and T. Zijlstra for technical support building the setup. Valuable discussions with D. Krug on the TNTI is also appreciated.

\bibliographystyle{elsarticle-harv} 
\bibliography{main}





\end{document}